\documentclass[a4paper,fleqn,usenatbib,useAMS]{mnras}
\usepackage[T1]{fontenc}
\usepackage{ae,aecompl}
\usepackage{graphicx}
\usepackage{amsmath}
\usepackage{amssymb}
\newcommand\blue[1]{\textcolor{blue}{\textbf{#1}}}

\title[A curious case of the pulsar GX~1+4]
{A curious case of the accretion-powered X-ray pulsar GX~1+4}

\author[Jaisawal et al.]
{Gaurava K. Jaisawal$^{1, 2}$\thanks{gaurava@space.dtu.dk}, Sachindra Naik$^2$, Shivangi Gupta$^2$, J{\'e}r{\^o}me Chenevez$^1$,
\newauthor Prahlad Epili$^2$\\
$^1$ National Space Institute, Technical University of Denmark, Elektrovej 327-328, DK-2800 Lyngby, Denmark\\ 
$^2$ Astronomy and Astrophysics Division, Physical Research Laboratory, Navrangapura, 
Ahmedabad - 380009, Gujarat, India\\}

\date{}

\pubyear{2017}

\begin{document}
\label{firstpage}
\pagerange{\pageref{firstpage}--\pageref{lastpage}}
\maketitle

\begin{abstract}

We present detailed spectral and timing studies using a {\it NuSTAR} observation 
of GX~1+4 in October 2015 during an intermediate intensity state. The measured 
spin period of 176.778~s is found to be one of the highest values since its discovery. 
In contrast to a broad sinusoidal-like pulse profile, a peculiar sharp peak is
observed in profiles below $\sim$25 keV. The profiles at higher energies are 
found to be significantly phase-shifted compared to the soft X-ray profiles. 
Broadband energy spectra of GX~1+4, obtained from {\it NuSTAR} and {\it Swift} 
observations, are described with various continuum models. Among 
these, a two component model consisting of a bremsstrahlung and a blackbody component 
is found to best-fit the phase-averaged and phase-resolved spectra. Physical 
models are also used to investigate the emission mechanism in the pulsar, 
which allows us to estimate the magnetic field strength  to be in 
$\sim$(5--10)$\times$10$^{12}$~G range. Phase-resolved spectroscopy of {\it NuSTAR} 
observation shows a strong blackbody emission component in a narrow pulse phase range. 
This component is interpreted as the origin of the peculiar peak in the pulse profiles 
below $\le$25~keV. The size of emitting region is calculated to be $\sim$400~m. 
The bremsstrahlung component is found to dominate in hard X-rays and explains 
the nature of simple profiles at high energies.
\end{abstract}

	
\section{Introduction}

Accretion powered X-ray pulsars constitute a large fraction of the brightest X-ray 
binaries in our Galaxy. These objects are highly magnetized rotating neutron 
stars that are powered by mass accretion from their optical companion. Depending 
on the characteristics of the donor star, mass transfer from the optical companion 
to the neutron star takes place through different mechanisms such as (i) Roche-lobe 
overflow (in low mass X-ray binaries), (ii) capture of stellar wind (in high mass 
X-ray binaries), and (iii) Be-disk accretion (in Be/X-ray binaries, a sub-class 
of high mass X-ray binaries; \citealt{Paul2011}). Irrespective of the accretion 
mechanisms, the magnetic field of the neutron star has a strong influence on the 
emission processes that leads towards a characteristic shape of X-ray continuum 
spectrum and beam geometry. Most of the high energy emission from the 
accretion-powered X-ray pulsars originates from the Comptonization of seed photons 
in the accretion column, formed on the surface of the neutron stars \citep{Becker2007}.  

GX~1+4 is an interesting accretion powered X-ray pulsar 
discovered in 1970 by a balloon-borne experiment at a spin period of about 2 minutes
\citep{Lewin1971}. It belongs to a rare type of symbiotic stars, the emerging class of 
X-ray binaries, consisting of bright objects that accrete matter from late 
type (K-M spectral class) giant companions \citep{Iben1996}. Generally a white dwarf 
is expected to be the accreting object in this type of systems, though a 
handful of systems with neutron stars as compact objects have been discovered in 
the last four decades, e.g. GX~1+4, 4U~1945+31, 4U~1700+24, Sct~X-1 and 
IGR~J16194-2810 ({\citealt{Corbet2008} and references therein). The 
optical counterpart of GX~1+4 is the bright infrared star V2116~Oph of 
M6~III spectral type, confirming the system to be a symbiotic binary 
(\citealt{Glass1973, Davidsen1977, Chakrabarty1997}). Assuming the 
neutron star mass as 1.35~M$_\odot$, the mass of the optical companion 
has been estimated to be 1.2~M$_\odot$ through the infrared radial velocity 
measurement method \citep{Hinkle2006}. The neutron star in this system 
is thought to accrete from dense stellar wind as the giant companion 
underfills its Roche-lobe (\citealt{Makishima1988, Hinkle2006}). 
The distance of the binary was estimated in the range of 3-15 kpc 
\citep{Chakrabarty1997}, and \citet{Hinkle2006} suggested it to be 
4.3~kpc. 

Since its discovery, the pulsar has exhibited an unusual pulse period 
history. It was first found to be spinning up during 1970s with 
the fastest period change rate ($\dot{P}$=-7.5$\times$10$^{-8}$ s~s$^{-1}$; 
\citealt{Makishima1988}) known among X-ray pulsars. Later, for 
an extended period in early 1980s, the pulsar became undetectable. 
When it reappeared, a torque reversal was observed and the source 
started showing spin-down behavior on similar magnitude as observed 
during the spinning-up era \citep{Makishima1988}. Based on the standard 
accretion torque theory, the torque-reversal event made it possible 
to predict the magnetic field of the neutron star to be $\sim$10$^{14}$~G 
(\citealt{Ghosh1979, Makishima1988, Dotani1989}). Alternative models such 
as accretion from retrograde disk are also suggested to resolve the unusual 
high value of  magnetic field and the rapid spin-down of the pulsar \citep{Makishima1988}. 
Using BATSE observations, an anti-correlation between torque and luminosity 
was detected and was explained by considering retrograde disk scenario 
\citep{Chakrabarty1997}. This retrograde disk concept also denied   
the need for a high magnetic field in GX~1+4. A tentative detection of a cyclotron 
absorption line at $\sim$34~keV has been reported from {\it BeppoSAX} and 
{\it INTEGRAL} observations, suggesting the magnetic field of the pulsar to 
be in the order of 10$^{12}$~G (\citealt{Naik2005, Rea2005, Ferrigno2007}). 
Though the retrograde disk hypothesis was supported by many authors 
(\citealt{Dotani1989, Chakrabarty1997}), recent long term spin-period 
evolution on top of sudden spin-up episodes can not be explained either 
by standard disk accretion model or retrograde disk theory, but rather 
through the quasi-spherical disk accretion on to the neutron star  
\citep{Gonzalez2012}. 

Using the pulse period variation during the spinning-up phase of 
pulsar, \citet{Cutler1986} proposed a periodicity of 304~d as the 
orbital period of the system. Detailed studies based on the infrared, 
optical and X-ray observations later established the orbital period 
to be 1161~d (\citealt{Hinkle2006, Ilkiewicz2017}). GX~1+4 is a 
peculiar source in many aspects. The pulse profiles of the pulsar 
show a characteristic sharp dip at medium and low intensity states 
which was interpreted as due to the eclipse of the emitting region by
the accretion column (\citealt{Giles2000, Galloway2000, Galloway2001}). 
The relationship between spin period history and the source luminosity 
is much more complicated than predicted by the standard accretion 
disk theory (\citealt{Ghosh1979, Chakrabarty1997, Gonzalez2012}). It 
occasionally shows a low flux state during which pulsations from the 
neutron star \blue{are} not observed (\citealt{Cui1997, Cui2004}). The absence 
of pulsation was interpreted as due to the onset of ``propeller effect'' 
and was detected for the first time in an accretion-powered X-ray pulsar 
\citep{Cui1997}.

The X-ray spectrum of GX~1+4 has been described by a cutoff power law model
during intermediate and high intensity states (10$^{-10}$--10$^{-9}$ erg 
cm$^{-2}$ s$^{-1}$ in 2-10 keV range; \citealt{Cui2004, Ferrigno2007}) or 
with a physical continuum model consisting of Comptonization of soft photons 
in hot plasma (\citealt{Galloway2000, Galloway2001, Naik2005}). Along with 
intrinsic absorption, strong iron emission lines at 6.5--7~keV with equivalent 
widths of 0.2--0.5 keV were detected in spectra obtained from {\it RXTE} and 
{\it BeppoSAX} observations. The pulsar also exhibits rare and irregular low 
intensity states (10$^{-11}$ erg cm$^{-2}$ s$^{-1}$ in 2-10 keV range) for 
durations varying from several days to month. During these low intensity states, the 
spectra were found to be highly variable (\citealt{Galloway2000,Rea2005,Naik2005}). 
The origin of the peculiar low 
intensity state is still ambiguous and is suggested to be associated with different 
mechanisms such as (i) propeller effect, (ii) accretion column eclipses, and 
(iii) obstruction through a thick accretion disk (\citealt{Cui2004, Galloway2000, 
Rea2005}). 

In the present work, comprehensive spectral and timing studies of GX~1+4 
have been carried out by using data from major broadband X-ray observatories 
such as {\it NuSTAR}, {\it Swift}, {\it RXTE} and {\it Suzaku}. The motivation 
of this study is to understand the energy and luminosity evolutions of the emission 
geometry of the pulsar by exploring pulse profiles at various intensity levels. A 
thorough investigation of the source spectrum is also carried out by using empirical 
and physical models. We have employed physical continuum models based on the 
thermal and bulk Comptonization processes in the accretion column (\citealt{Becker2007, 
Ferrigno2009, Farinelli2012}) to understand the column physics. In this paper, 
Section~2 describes the details of the observations and the procedures followed for 
analysing data from sevaral observatories. Our results and discussions are presented
in following sections.

\begin{figure*}
\centering
\includegraphics[height=5.3in, width=2.8in, angle=-90]{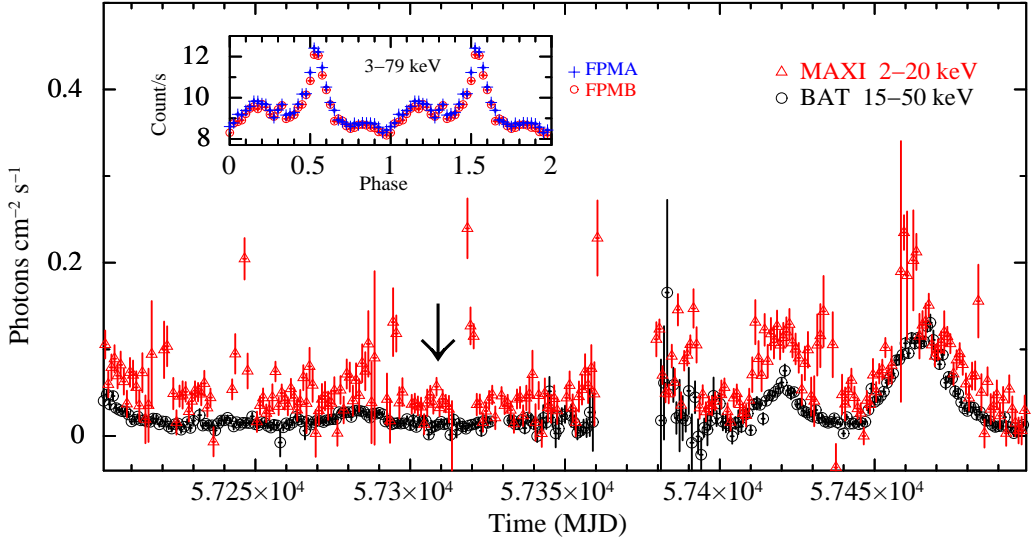}
\caption{Long term daily averaged monitoring light curve of GX~1+4 with {\it MAXI} (red) 
and {\it Swift}/BAT (black) in 2-20 and 15-50 keV energy ranges, respectively. The source 
flux obtained from BAT is scaled-up by a factor of three for better comparison with the 
data obtained from {\it MAXI}. The arrow mark in the figure indicates the date of the {\it NuSTAR} 
observation of the pulsar. Pulse profiles obtained from FPMA and FPMB detectors of 
{\it NuSTAR} are shown in the inset of the figure.} 
\label{fig1}
\end{figure*}

\section{Observations and Analysis}
 
GX~1+4 has been observed on several occasions at various intensity levels with X-ray 
observatories such as {\it Nuclear Spectroscopic Telescope Array (NuSTAR)}, {\it Swift}, 
{\it Suzaku} and {\it Rossi X-ray Timing Explorer (RXTE)}. We have used high quality 
{\it NuSTAR} and {\it Suzaku} data to explore the spectral characteristics and accretion 
column physics at distinct luminosity levels. Along with this, we have also analyzed a 
total of 143 pointing observations of the pulsar with the {\it RXTE} for an effective 
exposure of $\sim$551.53 ks performed during 1996 February to 2003 January to understand 
the evolution of beam function and spectral shape of the pulsar. A log of pointed 
observations used in this paper is given in Table~\ref{log} along with corresponding 
effective exposures. A detail description of observations and data analysis methods 
are presented in following section.

\begin{table}
\centering
\caption{Log of pointed observations of GX~1+4 with {\it RXTE}, {\it Suzaku}, 
{\it NuSTAR} and {\it Swift}.}
\begin{tabular}{ccccc}
\hline
\hline
{\it RXTE}     &No. of   &Time Range      &Expo. \\
Proposal ID    &Obs. IDs    &(MJD)           &(ks)\\
\hline
\
10133            &4    &50125.58--50126.53    &50.08 \\
10104            &23   &50130.12--50481.60    &135.26 \\
10103            &5    &50284.15--50284.85    &60.56 \\
10144            &1    &50464.13--50464.25    &10.28 \\
20170            &20   &50504.09--50588.29    &19.33 \\
60060            &40   &51974.70--52320.67    &120.9 \\
70064            &40   &52338.80--52593.68    &124.61 \\
70065            &8    &52390.87--52585.32    &28.95 \\
70425            &2    &52662.73--52670.41    &1.56 \\
\\
\hline
\hline
Observatory/	   &ObsID       &Start Date     &Expo.  \\
Instrument	       &	        &(MJD)				  &(ks) \\
\hline
{\it Suzaku}   	&405077010    &55471.28    	&99.8 \\ 
{\it NuSTAR}  	&30101040002  &57309.02     	&49.3 \\
{\it Swift}/XRT &00081653001  &57309.39 	&1.7 \\
 \hline
\hline
\end{tabular}
\label{log}
\end{table}

\subsection{{\it NuSTAR} and {\it Swift} Observations} 

{\it NuSTAR} is the first hard X-ray focusing observatory covering 3--79 keV energy 
range \citep{Harrison2013}. It was launched on 2012 June 13 in a low inclination 
Earth orbit with motivation to understand the enigma of X-ray cosmos. It carries 
two identical grazing angle focusing telescopes that reflect high energy photons 
below 79 keV to their respective focal plane modules, FPMA and FPMB. Each module 
comprises four 32$\times$32 Cadmium-Zinc-Telluride (CZT) detectors arranged in 
a plane. The total dimension of CZT detector is 20~mm$\times$20~mm with thickness 
of 2~mm. The focal length of each telescope is 10.15 m. The energy resolution of 
{\it NuSTAR} is 0.4 keV at 10 keV and 0.9~keV at 68 keV (FWHM), respectively. 

GX~1+4 was observed with {\it NuSTAR} on 2015 October 14 (MJD 57309.02) in a faint 
state. Long term monitoring light curves from {\it MAXI} and {\it Swift}/BAT are
shown in Fig.~\ref{fig1}. The arrow mark in the figure indicates the date of
{\it NuSTAR} observation of the pulsar for an effective exposure of $\sim$49.3~ks 
(Obs ID: 30101040002). The pulsar was also observed with the {\it Swift}/XRT 
(\citealt{Burrows2005}) on the same day with a net exposure of $\sim$1.7 ks. We followed standard 
procedures to analyze the {\it NuSTAR} data by using {\tt NUSTARDAS} software v1.4.1 
of {\tt HEASoft} version 6.16. We first reprocessed unfiltered events with the help 
of {\it nupipeline} task for both the detector units in the presence of recent 
calibration data base (CALDB) files. Further, science quality events, produced 
after the reprocessing, were utilized for extracting the barycentric corrected 
light curves, spectra, response matrices and effective area files by using {\it 
nuproducts} command. The source products were accumulated from a circular region 
of 120~arcsec around the central object for both the detectors. Moreover, 
background light curves and spectra were estimated in similar manner by 
considering a circular region of 120~arcsec away from the source.

{\it Swift}/XRT was operated in photon counting mode during the observation 
of GX~1+4. We have used data from this observation for spectral coverage in 
1-10~keV energy range. The unfiltered events were reprocessed by using {\tt 
XRTPIPELINE}. We noticed that the light curve obtained from cleaned data  
consisted of high count rate ($>$0.6) durations, indicating possible photon 
piled-up in XRT observation. We estimated the pile-up affected 
region across the source centre, as suggested by the instrument 
team\footnote{{http://www.swift.ac.uk/analysis/xrt/pileup.php}}. Accordingly, an 
annulus region with inner and outer radii of 10 and 60 arcsec was considered to 
extract source spectra from the cleaned XRT events using {\tt XSELECT} package. 
Background spectra were accumulated from a source free region in a similar manner. 
Response matrix and effective area files were also accumulated by following the 
standard procedure.

\subsection{{\it RXTE} Observations}

To investigate the properties of pulsar across a wide range of luminosity, we 
used publicly available {\it RXTE} observations carried out during 1996 February 
to 2003 January in our study (see Table~\ref{log}). Though most of these observations 
have been already used to understand the properties of GX~1+4 (\citealt{Cui1997, 
Galloway2000, Galloway2001, Cui2004, Serim2017}), we are motivated to examine long 
term spectral evolution of the pulsar to characterize its emission geometry.
The {\it RXTE} was a space-based observatory, launched in 1995 December by NASA. 
It worked for more than 16 years and 
extensively explored the X-ray sky. It had a broad-band energy coverage 
of 3--250 keV with two sets of onboard detectors such as (i) Proportional 
Counter array (PCA; \citealt{Jahoda1996}) and (ii) High Energy Timing Experiment 
(HEXTE; \citealt{Rothschild1998}). The PCA consisted of five Proportional Counter 
Units (PCUs) sensitive in 2--60 keV range with a total collecting area of 
$\sim$6500~cm$^{2}$. The HEXTE consisted of two clusters of detectors e.g., 
Cluster~A and Cluster~B, rocking orthogonally to each other for simultaneous 
measurement of source and background. Each cluster of HEXTE was a package of 
four NaI(Tl)/CsI(Na) phoswich scintillators, effective in the energy range of 
15--250 keV. The total collective area of HEXTE was $\sim$1600~cm$^{2}$. 

We followed standard procedures of data analysis as described in {\it RXTE cook 
book}\footnote{{https://heasarc.gsfc.nasa.gov/docs/xte/recipes/cook\_book.html}}. 
Source and background products were extracted by creating good time 
intervals by applying filter selections on all available PCUs. The 
source light curves were extracted in 2--60 keV range from Standard-1 
data at 0.125~s time resolution by using {\it saextrct} task of 
{\tt FTOOLS}. Corresponding background light curves and spectra 
were also generated from Standard-2 data by using background models 
provided by the instrument teams. For spectral studies, the source 
spectra were accumulated from Standard-2 data by using {\it saextrct} 
task. The response matrices were generated by using {\it pcarsp} task. 

\subsection{{\it Suzaku} Observation}

A {\it Suzaku} observation was also used in our study to understand the 
broad-band spectral properties of GX~1+4. {\it Suzaku}, the fifth Japanese 
X-ray satellite, was launched by Japanese Aerospace Exploration Agency in 
2005 July \citep{Mitsuda2007}. It carried two sets of major instruments such 
as X-ray Imaging Spectrometers (XISs; \citealt{Koyama2007}) and Hard X-ray 
Detectors (HXDs; \citealt{Takahashi2007}), providing a coverage of 0.2--600 
keV energy range. Four CCD instruments were the part of the XIS instruments, 
effectively operating in 0.2-12 keV range. The HXD unit of {\it Suzaku} 
consisted of two detectors such as HXD/PIN and HXD/GSO. The HXD/PIN consisted 
of silicon diode detectors, sensitive in 10--70 keV energy range, whereas 
HXD/GSO consisted of crystal scintillator detector working in the 40--600 keV 
energy range.

GX~1+4 was observed with {\it Suzaku} on 2010 October 2 for a total exposure 
of $\sim$195~ks (see Table~\ref{log}). Following the procedures described in 
{\it Suzaku ABC Guide}\footnote{{https://heasarc.gsfc.nasa.gov/docs/suzaku/analysis/abc/}}, 
source and background spectra from XIS-0, XIS-1, XIS-3, PIN and GSO instruments were 
accumulated and analysed to understand the broadband spectral characteristics of the pulsar.

\begin{figure*}
\centering
\includegraphics[height=6.1in, width=3.7in, angle=-90]{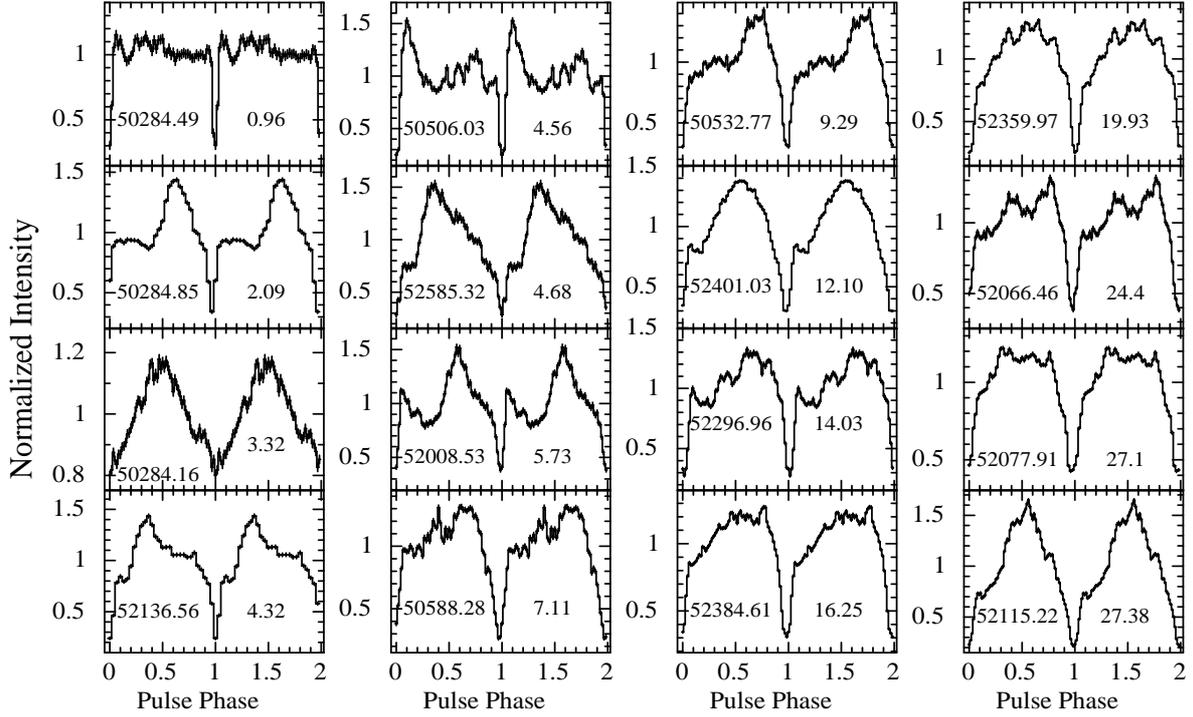}
\caption{Pulse profiles of the pulsar GX~1+4 obtained from {\it RXTE} observations at increasing 
intensity levels. These profiles were generated by folding the 2-60 keV light curves from 
PCA data at respective spin period. The numbers quoted in left and right side of each panel 
denote the beginning of the corresponding observation (in MJD) and the 3--30 keV unabsorbed 
flux (in unit of 10$^{-10}$ erg~cm$^{-2}$) s$^{-1}$, respectively. Two pulses are shown in 
each panel for clarity. The error bars represent 1$\sigma$ uncertainties.}
\label{fig2}
\end{figure*}
\begin{figure}
\centering
\includegraphics[height=3.35in, width=3.65in, angle=-90]{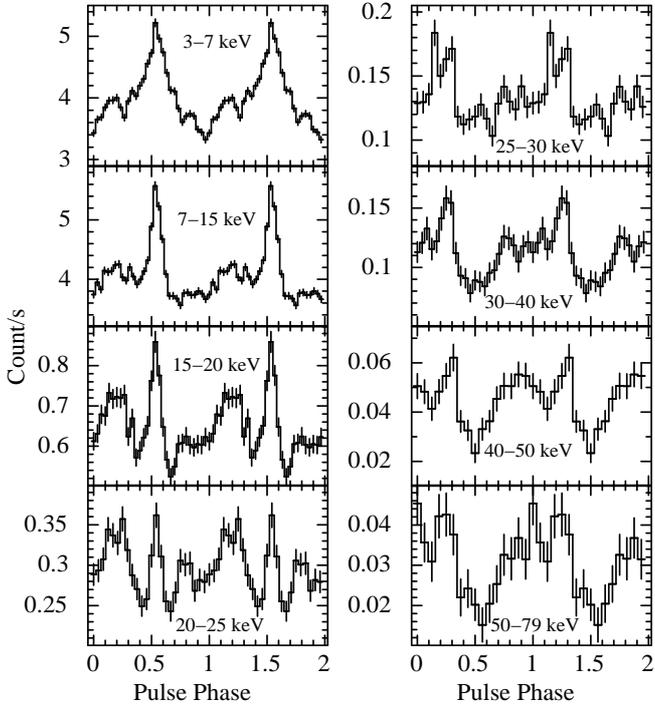}
\caption{Energy-resolved pulse profiles of GX~1+4 obtained from the {\it NuSTAR} 
observation in October 2015. A sharp peak along with plateau-like structure 
can be clearly seen in soft X-ray pulse profiles. Though the profile in 
$\ge$30~keV range is simple, a phase shift of $\pi$ with respect to the 
profile in 3--7 keV energy range is observed. The error bars represent 
1$\sigma$ uncertainties. Two pulses in each panel are shown for clarity.}
\label{fig3}
\end{figure}

\section{Timing analysis}

Following the procedures described in Section~2.1, source and background 
light curves were extracted from {\it NuSTAR} data in various energy bands 
at a time resolution of 0.1~s. Barycentric correction was applied on 
background subtracted light curves to incorporate the motion of earth 
and satellite to the barycenter of the solar system.  
We applied the $\chi^2$-maximization technique (\citealt{Leahy1987})
on the 3--79 keV barycentric corrected light curve to estimate the spin 
period of the pulsar. Using this technique, X-ray pulsations at 176.778(6)~s 
were detected in the pulsar light curve. We also confirm the 
estimated spin period by using another software package e.g. {\it HENDRICS - 
High ENergy Data Reduction Interface from the Command Shell} 
(\citealt{Bachetti2015}). This package performs a pulsation search by 
following different approaches such as epoch-folding and the Z${^2_n}$ 
statistics \citep{Buccheri1983} on clean events. Using both methods, a best fitted period 
at 176.778~s was detected in the {\it NuSTAR} data of the pulsar. Based on 
the agreement on the estimated values of periodicity through independent methods,
the spin period of GX~1+4 was considered to be 176.778(6)~s. This is one 
of the highest value of spin period of the pulsar since its discovery, 
indicating a continuous spin-down trend after torque reversal in early 1980s.

Pulse profiles were generated by folding the 3--79 keV light curves from 
the FPMA and FPMB detectors of {\it NuSTAR} at the measured pulsation period
and are shown in the inset in Fig.~\ref{fig1}. A plateau like feature in
the pulse profile at $<$0.4 phase, followed by a peculiar sharp peak in 0.5--0.6 
phase range is seen (Fig.~\ref{fig1}). Such type of unusual features have not 
been detected in the pulse profiles of GX~1+4 obtained from {\it RXTE}, {\it BeppoSAX}, 
{\it INTEGRAL} and {\it Suzaku} observations (\citealt{Cui2004, Naik2005, 
Ferrigno2007, Yoshida2017}).  The effect of spin period derivative ($\dot{P}$=
10$^{-7}$~s~s$^{-1}$; \citealt{Gonzalez2012}) was considered while checking
the shape of pulse profile of the pulsar from {\it NuSTAR} observation.
Indifferent shape of the profile confirmed the presence of a peculiar emission 
geometry during the {\it NuSTAR} observation.

\subsection{Luminosity dependence of pulse profiles from {\it RXTE} observations}

To thoroughly investigate the long term pulse profile evolution of the 
pulsar and compare the same from the {\it NuSTAR} observation, a detailed 
timing studies were performed using all available {\it RXTE} observations of 
GX~1+4 during the period of 1996 to 2003. These pointed observations were carried 
out at various flux levels of the pulsar, providing a unique opportunity to probe 
the energy and luminosity evolutions of the beam geometry and column physics. For 
this, the 2--60 keV light curves, generated from Standard-1 data, were used to 
obtain the pulse profiles from each of the observations.  This was done by folding 
the light curves at their respective estimated spin periods in between 123.5 to 139~s. 
The epochs were manually adjusted to align the profiles such that the primary dip 
(minimum) was always at phase 0. Representing the pulse profiles in such a way can
make it suitable to investigate the evolution of emission geometry during various 
pointings.  Some of the representative
pulse profiles obtained from the {\it RXTE} observations are shown in Fig.~\ref{fig2} 
in increasing order of source flux. The date of corresponding observations (MJD) are 
also quoted in the pulse profiles. The shape of pulse profiles appeared to be broad. 
Apart from the broad shape, the profiles did not show any long term luminosity 
dependence or any systematic change. However, the evolution of beam geometry or 
pulse profile can be easily traced among successive {\it RXTE} observations.  A 
sharp peak like profile as seen during {\it NuSTAR} observation (not exactly the 
same) was also noticed in the figure at MJD 50284.85 and 52008.53 (i.e. second and 
seventh panels of Fig.~\ref{fig2}). Moreover, the detailed energy evolution of 
these profiles was found to be simple and consistent with a broad sinusoidal 
like profile at higher energies.

\subsection{Energy dependence of pulse profiles from {\it NuSTAR} observation}

Evolution of pulse profiles with energy was investigated out by folding 
the background subtracted energy resolved light curves obtained from the
{\it NuSTAR} observation, with the estimated spin period of the pulsar. 
These pulse profiles are shown in Fig.~\ref{fig3}. The profiles were found
to be strongly energy dependent in contrast to the profiles seen during the 
{\it RXTE} era. A broad peak in 0.3--0.7 phase range of 3--7 keV profile  
evolved to a very narrow peak with increasing energy up to $\le$25~keV 
which then completely disappeared from the profiles at higher energies. 
Pulse profiles at $\ge$30~keV are found to be singly peaked and $\sim$180 
degree phase shifted with respect to the 3--7 keV pulse profile. From the 
evolution of the pulse profiles with energy, it is apparent that two different  
components are contributing to the pulsar emission during the {\it NuSTAR} 
observation. One of the component, dominating below 25 keV, is thought to be
originated from a narrow phase range ($\sim$0.4--0.6) of the pulsar whereas 
the other component contributed significantly in the hard X-ray range. Based 
on our study, it is clear that the pulse profile of the pulsar during the 
{\it NuSTAR} observation is unique and seen for the first time in the history 
of GX~1+4. To probe the origin of the observed peculiar structures and its 
distinct emission components, phase-averaged and phase-resolved spectroscopy 
are performed and described in the next section.

\begin{table*}
\centering
\caption{Best-fitting spectral parameters (90\% errors) obtained from the {\it NuSTAR} and 
{\it Swift}/XRT observations of GX~1+4. The fitting models consist of (i) a high-energy 
cutoff power law with blackbody, (ii) cutoff power law model with blackbody, (iii) CompTT with bremsstrahlung (or blackbody) component, (iv) bremsstrahlung with blackbody component and (v) COMPMAG2 model along with photoelectric absorption component and two Gaussian components for iron lines.}
\begin{tabular}{ |l | ccccccc}
\hline 
Parameters                      &  \multicolumn{7}{c}{Spectral Models}    \\ 
&  \\\cline{2-8}  \\
                                &HECut+BB     &Cutoff+BB     &CompTT+Br     &CompTT+BB	&NPEX+BB	&Br+BB     &COMPMAG2	 \\
\hline
N$_{H}$$^a$                    &2.55$\pm$0.32         &2.48$\pm$0.22     &3.62$\pm$0.6	   &0.91$\pm$0.23       	&2.18$\pm$0.24 		&2.66$\pm$0.15 	 &1.2$\pm$0.1 \\
Photon index          &1.32$\pm$0.04       &1.2$\pm$0.04      &--		      &--		&1.09$\pm$0.08 		 &--	\\
E$_{cut}$ (keV)	                &7.2$\pm$0.8         &33.6$\pm$2.4        &--      	    &--                  &25.73$\pm$4.5		 &--	&--\\
E$_{fold}$ (keV)	      &37.7$\pm$3.1              &-- 		 &--                 &--                    	&-- 		 &--	 &--\\
BB temp. (keV)       		&2.01$\pm$0.15	    &2.04$\pm$0.07 		&--        &2.8$\pm$0.1         &1.94$\pm$0.09		&2.16$\pm$0.05  &--	\\        
Bremss temp. (keV)                &--                    &--             &1.7$\pm$0.5         &--                 &--  			&42.91$\pm$1.31		&--\\
CompTT T$_0$ (keV)                &--                   &--             &1.5$\pm$0.1         &0.95$\pm$0.05 	&--			&-- 	&--\\
CompTT $\tau$                     &--                   &--             &2.5$\pm$0.1        &3.3$\pm$0.2            &--			&-- 	&--\\
CompTT kT (keV)                   &--                   &--             &14.3$\pm$0.8       &12.2$\pm$0.6           &--			&-- 	&--\\
\\
COMPMAG kT$_{bb}$ (keV)                  &--                  &--              &--                &--   &--  &--  &1.37$\pm$0.02 \\
COMPMAG kT$_{e}$ (keV)   &--                  &--              &--                &--   &--  &--  &0.81$^{+1.2}_{-0.81}$ \\
COMPMAG $\tau$                   &--                  &--              &--                &--   &--  &--  &0.31$\pm$0.01 \\
Column radius (m)                   &--                  &--              &--                &--   &--  &--  &283$\pm$51 \\
\\
{\it Emission lines } \\
Line energy (keV)              &6.36$\pm$0.02       &6.36$\pm$0.02      &6.36$\pm$0.01     &6.36$\pm$0.01 	&6.36$\pm$0.01 		&6.36$\pm$0.02		&6.36$\pm$0.01\\
Eq. width  (eV)               &99.8$\pm$9.2	    &102.3$\pm$9.1      &93.2$\pm$6.9      &115.2$\pm$10.4      &102.2$\pm$7.5 		&105.6$\pm$8.9	&94.8$\pm$9.8 \\
Line energy (keV)              &6.94$\pm$0.16       &6.97$\pm$0.08      &6.99$\pm$0.1     &6.96$\pm$0.06 	&6.96$\pm$0.07 		&6.99$\pm$0.07	 &6.96$\pm$0.08\\
Eq. width (eV)              &17.4$\pm$10.1	   &22.1$\pm$7.2        &16.7$\pm$6.2     &29.3$\pm$6.8         &22$\pm$7		&23.9$\pm$6.1  &19.3$\pm$8.4 \\
\\
{\it Component Flux}$^b$	
\\
Power-law flux	        &5.95$\pm$0.6	    &5.8$\pm$0.5	&--		   &--				&5.82$\pm$0.5		&--	&--\\
Blackbody flux		&0.33$\pm$0.05	    &0.49$\pm$0.05	&--			&0.67$\pm$0.07		&0.44$\pm$0.03		&0.44$\pm$0.05	&--\\
CompTT flux			&--		&--		&6.0$\pm$0.4		&5.4$\pm$0.3		&--			&--		&--\\
Bremsstrahlung flux 		&--		&--	  &0.37$^{+0.22}_{-0.12}$	   &--			&--			&5.9$\pm$0.1		&--\\
\\
{\it Source flux}$^b$
 \\
Flux (3-10 keV)   		&1.96$\pm$0.07      &2.05$\pm$0.17       &2.08$\pm$0.36       &1.84$\pm$0.11      &1.94$\pm$0.04   	&1.98$\pm$0.05     &1.86$\pm$0.1\\
Flux (10-70 keV)  		&4.35$\pm$0.22      &4.35$\pm$0.42       &4.35$\pm$0.31       &4.31$\pm$0.26      &4.36$\pm$0.4    	 &4.37$\pm$0.08  &4.34$\pm$0.15\\ 
\\
Reduced $\chi^2$ (\it d.o.f)         &0.97 (795)          &0.98 (795)          &0.99 (795)        &1.07 (794)       &0.97 (794)     	  &0.99 (796)   &1.02 (794) \\
\hline
\end{tabular}
\flushleft
$^a$ : Equivalent hydrogen column density (in 10$^{22}$ atoms cm$^{-2}$ unit),
$^b$ : Unabsorbed flux in unit of 10$^{-10}$ ~ergs~cm$^{-2}$~s$^{-1}$. 
\label{table2}
\end{table*}

\section{Spectral analysis}

\subsection {Phase-averaged spectroscopy with {\it NuSTAR} and {\it Swift}/XRT} 

A detailed investigation on the presence of different emission components 
in GX~1+4 was carried out by using {\it NuSTAR} and {\it Swift}/XRT 
observations in a faint state in 2015 October. For broad-band 
spectral analysis, source and background spectra were accumulated by 
following the procedures described in earlier section. With appropriate 
background, response matrices and effective area files, the 1-79 keV 
energy spectra of GX~1+4 were fitted with several continuum models 
in {\tt XSPEC} package (ver. 12.8.2). In our simultaneous spectral fitting, 
all the model parameters were tied together for spectra obtained from FPMA, 
FPMB and XRT detectors whereas the relative instrument normalizations were 
kept free.


\begin{figure}
\centering
\includegraphics[height=3.2in, width=5.5in, angle=-90]{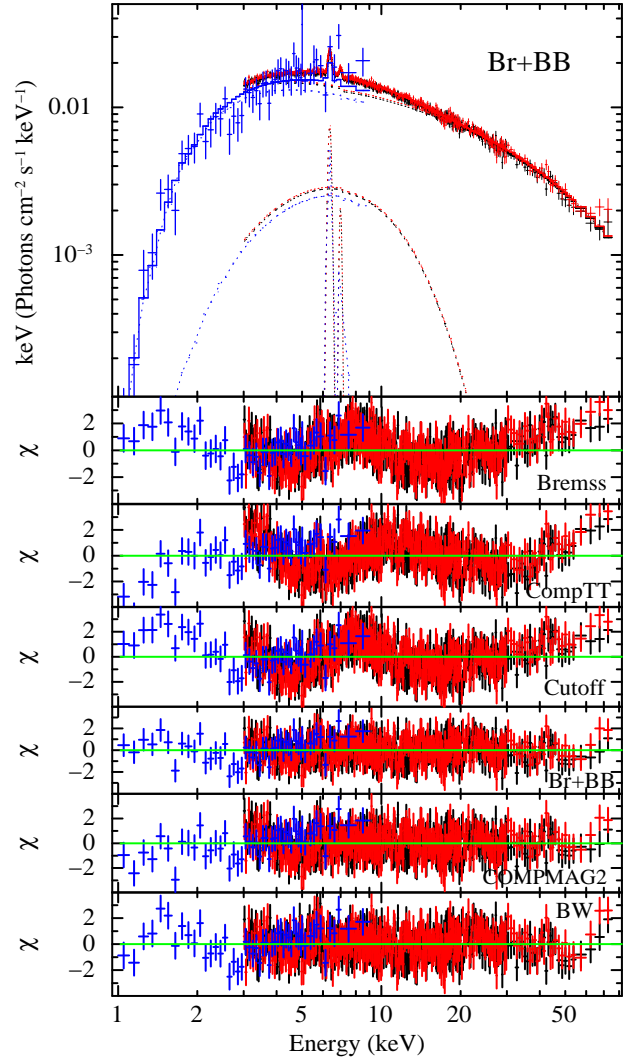}
\caption{Energy spectra of GX~1+4 in 1-79~keV range obtained from {\it Swift}/XRT and 
FPMA and FPMB detectors of {\it NuSTAR} from October 2015 observations along with the 
best-fit model comprising a bremsstrahlung and blackbody (Br+BB) model and two iron 
emission lines on top panel. Second, third and fourth panels show the contributions of the residuals 
to $\chi^{2}$ when the pulsar continuum was fitted with Bremsstrahlung, CompTT and Cutoff 
power law models, respectively. In all these panels, a broad excess in 4-20 keV range is 
clearly visible. The fifth panel shows the residuals for the model consisting of 
Bremsstrahlung and blackbody components. The sixth and seventh panels show the 
residuals obtained after fitting the 1-79 keV spectra with physical models such 
as COMPMAG2 and BW continuum models, respectively. Broad excess observed in 4-20 keV 
range is absent in the residuals (sixth and seventh panels) while fitting the data with 
COMPMAG2 and BW physical models.}   
\label{sp}
\end{figure}

Standard continuum models for accretion powered X-ray pulsars such 
as high energy cutoff power law (HECut), Negative and Positive Exponential 
Cutoff (NPEX), Fermi-Dirac cutoff power law, cutoff power law and 
Comptonization (CompTT) models along with photoelectric absorption 
component were applied to fit the energy spectrum of GX~1+4. However, 
none of the above models was suitable to fit the 1-79 keV spectrum 
obtained from {\it NuSTAR} and {\it Swift}/XRT 
observations. Presence of a broad excess in 4--20 keV energy range 
(third and fourth panels of Fig.~\ref{sp}) caused the simultaneous 
spectral fitting with above models unacceptable. Usually, the accretion 
powered X-ray pulsars show an excess in soft X-rays, known as soft excess 
in their spectra \citep{Hickox2004}. The soft excess feature is interpreted 
as due to the reprocessing of hard X-ray photons in the surrounding regions 
of the neutron star and described by a thermal emission component. Based on similar 
analogy, we attempted to express the broad excess observed in GX~1+4 
with a blackbody or bremsstrahlung component. Addition of a blackbody 
component to the above continuum models provided an acceptable fit 
by bringing down the reduced $\chi^2$ value from $\ge$ 1.5 to $\sim$1 
in each case. The blackbody temperature obtained from the fitting was 
found to be in the range of 2--3 keV (see Table~\ref{table2}). Such high 
temperature thermal component is difficult to understand in the soft 
excess scenario. Therefore, we presumed that the excess seen during the 
{\it NuSTAR} and {\it Swift}/XRT observations could be associated with 
the continuum from the accretion column and can be explained by a 
composite or physical model. A similar type of continuum i.e. cutoff 
power law with blackbody (Cutoff+BB) was used earlier to explain 
the pulsar spectra during high and intermediate states with {\it Suzaku}, 
{\it RXTE} and {\it BeppoSAX} \citep{Yoshida2017}.   

Instead of the blackbody component, a bremsstrahlung component was also tried 
with above ad-hoc models for understanding the nature of excess during {\it NuSTAR} 
observation. We found that the CompTT model with bremsstrahlung component 
(CompTT+Br) can fit the continuum as well. Apart from the 
interesting spectral shape, two iron emission lines at $\sim$6.4 and 6.9~keV are 
also detected in the spectrum of GX~1+4 during 2015 October observation.
We did not find any signature of cyclotron absorption feature in the 1-79 keV 
range. This is in contrast to the previous tentative detections of a cyclotron 
line at $\sim$34~keV from {\it BeppoSAX} and {\it INTEGRAL} observations 
(\citealt{Naik2005, Rea2005, Ferrigno2007}). Moreover the source flux during 
the {\it NuSTAR} observation was found to be lower compared to these earlier observations. 
With {\it NuSTAR}, the 3-10 keV unabsorbed flux was estimated to be 
$\sim$2$\times$10$^{-10}$ erg~cm$^{-2}$~s$^{-1}$, indicating an intermediate 
intensity state of the pulsar during the observation.  

While exploring the suitable spectral model for GX~1+4, we noticed that 
a two-component thermal model i.e. bremsstrahlung and blackbody components 
(Br+BB) along with interstellar absorption and two Gaussian functions for 
iron emission lines at $\sim$6.4 and 6.9 keV also fitted the spectra
well. The equivalent widths of the 6.4 and 6.9~keV lines were estimated 
to be $\sim$106~eV and 24~eV, respectively.
The bremsstrahlung component in this model was found to describe 
the pulsar continuum, producing a broad-excess in 4-20 keV range which 
was explained by a thermal blackbody component (second and fifth panels 
of Fig.~\ref{sp}). The Br+BB model was found statistically comparable to 
other models used to describe GX~1+4 spectrum from {\it NuSTAR} and 
{\it Swift}/XRT observations. Spectral parameters obtained from all the 
models are presented in Table~\ref{table2}.  The equivalent hydrogen 
column density is found to be variable in the range of 1--4$\times$10$^{22}$ 
cm$^{-2}$ depending on the continuum model. This is higher 
than the estimated value of Galactic absorption in the direction of the
source ($\sim$3$\times$10$^{21}$ cm$^{-2}$). Therefore, the presence of 
an additional absorber close to the neutron star is expected. However, pulse 
phase dependency of column density can not be explored in the present study 
due to limiting coverage  in soft X-ray ranges $\le$3~keV. 
The energy spectra of GX~1+4 for the best fitting Br+BB model is presented in 
the first panel of Fig.~\ref{sp}. Second, third and fourth panels of the figure 
show the residuals obtained after fitting the spectra with bremsstrahlung, CompTT 
and cutoff power law models, respectively. The fifth  panel indicates 
the residuals obtained after adding a blackbody component with the 
bremsstrahlung model.

\subsection {Phase-averaged spectroscopy with physical models}

To understand the properties of accretion column, we attempted to 
fit the pulsar spectra obtained from {\it NuSTAR} and {\it Swift}/XRT 
observations with COMPMAG \citep{Farinelli2012} and Becker and Wolff 
(BW; \citealt{Becker2007}) models. These continuum models are derived 
by considering slightly different assumptions on the accretion geometry, 
emission processes, velocity profile and also use different methods to 
solve the radiative transfer equation of photons through the accretion 
column.  

The COMPMAG model can successfully produce the spectral shape of faint 
accreting X-ray pulsars and supergiant X-ray transients. This model 
computes the bulk and thermal Comptonization of blackbody seed photons 
during cylindrical accretion on to the poles of the magnetized neutron 
star, assuming two different velocity profiles characterized by $\eta$ 
and terminal velocity $\beta$ near the surface \citep{Farinelli2012}. 
We have used this model in our study with reasonable parameter values 
consistent to low luminosity pulsars i.e. free fall velocity profile 
$\eta$=0.5, terminal velocity $\beta$=0.5 and the albedo of neutron 
star surface (A) to be 1 \citep{Farinelli2012}. The pulsar spectra 
when fitted with the COMPMAG model using general velocity profile 
(hereafter COMPMAG1) produced a broad excess in the residuals similar 
to that obtained while fitting the data with the empirical models. 
Addition of a blackbody component at $\sim$1.8~keV temperature yielded 
an acceptable fit. We have also fitted the spectrum of GX~1+4 by using 
a specific velocity profile of COMPMAG (hereafter COMPMAG2) that is 
linearly dependent on the optical depth ($\beta(\tau)$$\propto$$\tau$). 
The spectra fitted with the latter velocity profile resolved the peculiar 
broad excess and described continuum well with a reduced~$\chi^2$ close 
to 1. The residual obtained from COMPMAG2 is shown in the sixth panel of 
Fig.~\ref{sp}. Corresponding parameters obtained from data fitting are presented 
in Table~\ref{table2}. Based on these results, we expect that the broad 
feature (with ad-hoc and COMPMAG1 models) is intrinsically associated 
with the emission from accretion column or region located close to the 
neutron star. Moreover, the effect of bulk Comptonization of blackbody seed photons 
was found to dominate the continuum as the low value of electron plasma 
temperature was detected at high $\beta$.

The BW model was also used to explore the spectrum of GX~1+4. This is a 
physics based complex model that considers the effects of thermal and bulk 
Comptonization of seed photons, originated from bremsstrahlung, blackbody, 
and cyclotron emissions with accreting plasma in the column \citep{Becker2007}. 
It is best suitable for explaining the spectra of bright pulsars such as 4U~0115+63 
\citep{Ferrigno2009} and EXO~2030+375 \citep{Epili2017} in the presence of a
radiation dominating shock in the accretion column. For a canonical neutron 
star, BW model consists of six free parameters i.e. the ratio between bulk and 
thermal Comptonization $\delta$, a dimension-less parameter related to photon 
escape time $\xi$, magnetic field $B$, mass accretion rate $\dot{M}$, electron 
plasma temperature $T_e$, and accretion column radius $r_0$. 

We found that the pulsar spectrum in 1-79 keV range can be 
represented by the BW model. This model was able to resolve the puzzling 
4-20 keV broad excess as seen while fitting the data with the traditional and 
COMPMAG1 models (see Fig.~\ref{sp}). This also explains the presence of excess 
emission over the continuum from the accretion column or regions close to the 
pulsar. For spectral fitting with the BW model, the mass 
accretion rate was calculated by using the values of source flux estimated 
from empirical models and the distance of the source as 4.3 kpc 
\citep{Hinkle2006}. It is important to mention that the source distance 
is not well constrained because of the uncertainties in estimating the 
binary parameters. The uncertainties in source distance, therefore, affects the 
estimation of column radius which is strongly dependent on mass accretion rate 
\citep{Becker2007}. We have employed the BW model by considering a source distance 
of 4.3 and 10~kpc to compare the effect of mass accretion on other spectral 
parameters. We found that the parameters obtained at both the distances  
are nearly equal (within errors) except the column radius and normalization 
constant of the BW model. While fitting, the mass accretion rate $\dot{M}$ was 
kept frozen at respective value of distances. As the column radius strongly depends 
on the accretion rate, we also fixed the value of $r_0$ after best fitting of 
the model to get a better constraint on other parameters, as suggested by 
\citet{Ferrigno2009}. The column radius was found to be $\sim$5 and 10.5~m 
at source distances of 4.3 and 10~kpc, respectively. This model also provides 
opportunity to constrain the magnetic field of the pulsar along with other 
physical parameters. In our study, the value of magnetic field for GX~1+4 
was found to be $\sim$5.6$\times$10$^{12}$~G using {\it NuSTAR} and XRT data. 
We noticed that the value of magnetic field obtained from this model is   
insensitive to upper bound of the parameter. Therefore, only a lower value 
of field strength is reported in Table~\ref{table3}. Residual obtained from 
the spectral fitting with the BW model is shown in the seventh panel of 
Fig.~\ref{sp}.


\begin{figure}
\centering
\includegraphics[height=3.25in, width=2.5in, angle=-90]{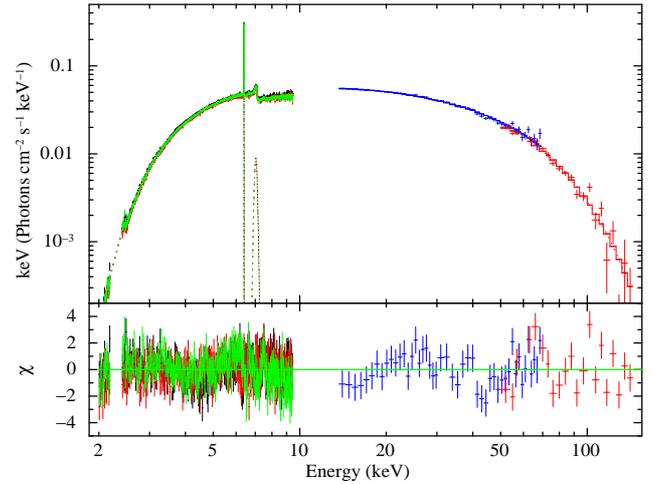}
\caption{Energy spectrum of GX~1+4 in 2-150~keV range obtained from XIS-0, 
XIS-1, XIS-3, HXD/PIN and HXD/GSO instruments of {\it Suzaku} during October 
2010 observation along with the best-fit BW model and two iron emission 
lines. The bottom panel shows the contributions of the residuals to $\chi^{2}$.}   
\label{sp2}
\end{figure}

\begin{table}
\centering
\caption{Parameters obtained from the spectral fitting of the data obtained from 
the {\it NuSTAR} and {\it Suzaku} observations of the pulsar during intermediate 
intensity levels with the BW model.}
\begin{tabular}{lcr}
\hline
\hline
 Parameters                        &{\it NuSTAR}	 &{\it Suzaku} \\
\hline

N$_{H}$$^a$                        &0.78$\pm$0.1	 &13.93$\pm$0.07 	\\
$\xi$				   &1.42$\pm$0.02	 &4.9$^{+3.2}_{-1.3}$	\\
$\delta$			   &12.86$\pm$0.87	 &0.55$\pm$0.25	    	\\
Electron temp. T$_e$ (keV)	   &4.3$\pm$0.1	   	 &11.7$^{+1.2}_{-0.6}$	\\
Column radius $r_0$ (m)		   &10.52	         &48.22			\\
Magnetic field B (10$^{12}$ G)	   &$>$4.6	         &8.2$^{+1.8}_{-0.6}$	\\
Accretion rate$^b$ (10$^{17}$ g/s) &0.4                  &2.87       		\\
\\
Source flux (3-70 keV)$^c$	   &6.18$\pm$0.24    	 &38.15$\pm$10.43	\\
Reduced $\chi^2$ (\it d.o.f)       &1.02 (799)       	 &1.06 (841)       	\\
\hline
\hline
\end{tabular}
\flushleft
$^a$ : Equivalent hydrogen column density (in 10$^{22}$ atoms cm$^{-2}$ unit), 
$^b$ : Mass accretion rate is calculated by considering a distance of 10 kpc, 
$^c$ : Unabsorbed flux (in unit of 10$^{-10}$ erg cm$^{-2}$ s$^{-1}$). 
\label{table3}
\end{table}

In order to constrain the magnetic field and understand the physics of accretion
column at different intensity levels of GX~1+4, we used data from a {\it Suzaku} 
observation of the pulsar in our analysis. This observation was carried out in 2010 
October at an intermediate intensity level (3--10 keV flux $\sim$7$\times$10$^{-10}$ 
erg cm$^{-2}$ s$^{-1}$; also see \citealt{Yoshida2017}). The phase-averaged spectra 
obtained from XIS-0, XIS-1, XIS-3, PIN and GSO detectors of {\it Suzaku} observation 
were fitted with the BW model in 2-150 keV energy range. As the pulsar was 
relatively bright, a systematic error of 1\% was added to the data from XISs to 
incorporate the cross calibration among front and back illuminated CCDs, as  
suggested in \citet{Epili2016} and references therein. Considering a source 
distance of 4.3 and 10~kpc, the parameters obtained from the fitting the data 
with the BW model were found to be consistent (within errors) with the earlier 
values, though the column radii were estimated to be $\sim$20.2 and 48.2~m, 
respectively. These values were relatively small compared to the column radii 
obtained from COMPMAG2 model. The estimated strength of magnetic field from the 
{\it Suzaku} observation was in the range of $\sim$(7--10)$\times$10$^{12}$~G. This 
value was found to be marginally higher compared to that computed from the {\it NuSTAR} 
and {\it Swift} observations. All the spectral parameters obtained from fitting {\it 
Suzaku} data with the BW model are presented in Table~\ref{table3}. We have 
described the implication of these results in the discussion section.

\begin{figure*}
\centerline{\includegraphics[angle=-90,width=17.cm]{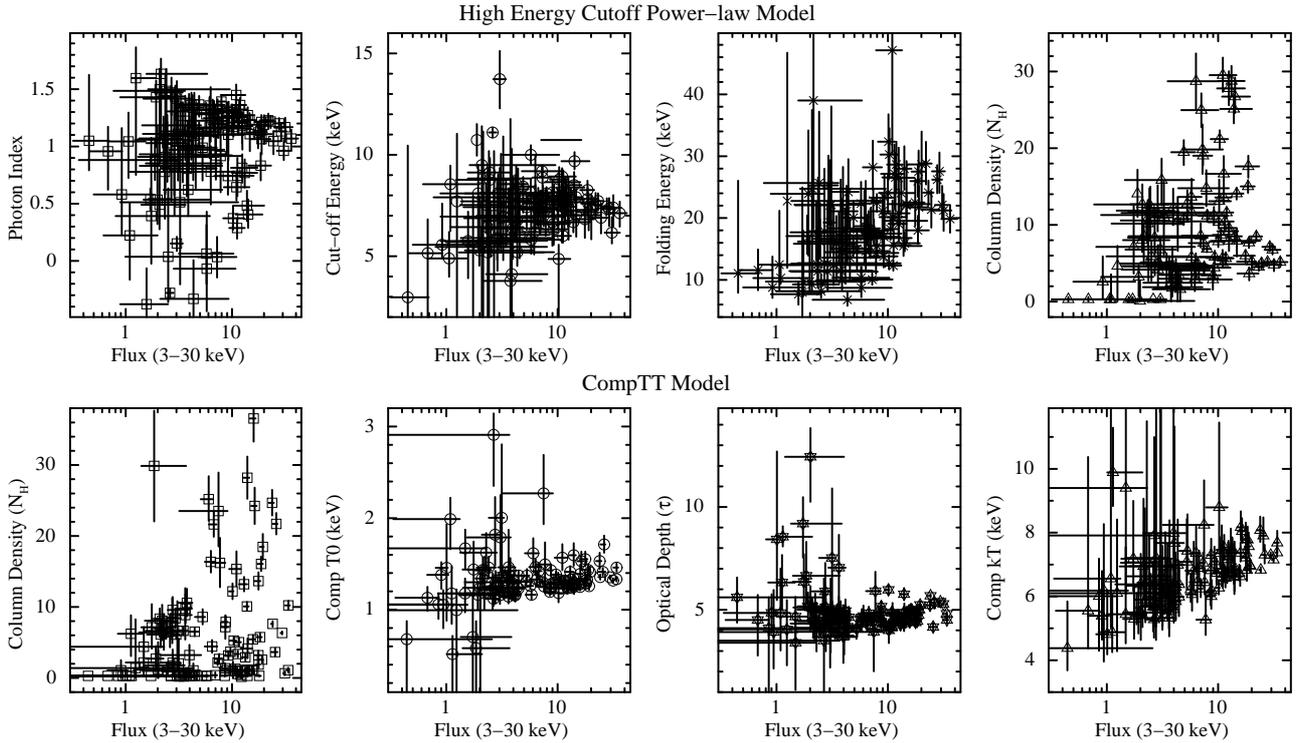}}
\caption{Variation of spectral parameters with source flux, obtained from 
phase-averaged spectroscopy of data obtained from the {\it RXTE} observations 
of GX~1+4 during February 1996 to  January 2003 by using high energy cutoff power 
law (HECut) and CompTT models. Top panels show the values of photon index, cutoff 
energy, folding energy and column density with respect to the unabsorbed 3-30 keV 
flux (in units of 10$^{-10}$~erg cm$^{-2}$ s$^{-1}$) from left to right, respectively, 
for the HECut model. For the CompTT model, variation of parameters such as column density, 
input soft temperature (T$_0$), optical depth and plasma temperature ($kT$) with 
respect to unabsorbed 3-30 keV flux are shown in bottom panels (from left to 
right, respectively). The errors in the figure are estimated for 90\% confidence 
level.}
\label{rxte-spec}
\end{figure*}

\subsection {Intensity dependent spectral studies of GX~1+4 with {\it RXTE}}

It has been found that the phase-averaged spectrum of GX~1+4 can be described 
with several continuum models such as CompTT, Cutoff+BB, HECut models 
(\citealt{Galloway2001, Naik2005, Ferrigno2007, Yoshida2017, Serim2017}). 
To investigate the spectral evolution of the pulsar with luminosity and 
understand the cause of the peculiar features observed during the {\it NuSTAR} 
and {\it Swift}/XRT observations (present work), all the {\it RXTE} observations 
of the pulsar during 1996 February to 2003 January have been analysed. In the 
beginning, we attempted to fit the 3--30 keV {\it RXTE}/PCA spectra with 
several models, as described in Section~4.1. While fitting, a systematic 
error of 0.5\% was also added to the spectra. We found that a HECut or  
CompTT model fits well to the pulsar continuum across a wide range of source 
intensity. In a few cases, a partial absorption component was also required to
fit the data well. While fitting the data during intermediate and high flux 
states, the pulsar continuum was found relatively complicated. 
\citet{Yoshida2017} used Cutoff+BB continuum model to describe pulsar spectrum 
at such intensity level. Based on the results obtained from present comparative 
analysis, we suggest that HECut or CompTT can describe the pulsar continuum 
better than cutoff power law model. \citet{Serim2017} also reported similar 
results by using the {\it RXTE} observations of GX~1+4 during 2001--2003. 
Variation of spectral parameters obtained from fitting data with the HECut 
and CompTT models with the absorption corrected flux in 3--30~keV range are 
shown in top and bottom panels of Fig.~\ref{rxte-spec}, respectively. 

For HECut model, parameters such as power-law photon index, cutoff energy, 
folding energy and galactic absorption column density are shown in the top 
panels of the figure. The power-law photon index was found to remain comparable
(within errors) for a wide flux range of 4--40 $\times$10$^{-10}$ erg cm$^{-2}$ 
s$^{-1}$ although moderate variations are seen for a few observations. The cutoff 
energy did not show any significant variability over the wide flux range. However, 
the folding energy showed a positive correlation with source flux. Though the 
absorption column density, obtained from HECut and CompTT models, did not show 
any systematic pattern, its values were high during medium and high intensity 
states of the pulsar. Spectral parameters from CompTT model such as optical 
depth and input soft photon temperature (T$_0$) were found to be relatively 
constant with the source flux. As in case of folding energy (HECut model), 
a positive correlation was also seen between plasma temperature and source 
flux.   
 
To compare the results obtained from {\it NuSTAR} and {\it Swift} observations 
with the {\it RXTE} observations of the pulsar, we attempted to fit the {\it RXTE} 
spectra with Br+BB model. However, we found that this model did not fit the data from 
any of the {\it RXTE} observations. Shape of the pulse profile obtained from 
2015 October {\it NuSTAR} observation was also unique and entirely different
from the profiles obtained from all the available {\it RXTE} observations of
the pulsar. This indicates that the 2015 October {\it NuSTAR} observation of
GX~1+4 was peculiar and different from all other observations.

\subsection {Phase-resolved spectroscopy during a peculiar state with {\it NuSTAR}}

Based on the results obtained from a detailed spectral and timing studies of GX~1+4 
by using the {\it RXTE} and {\it Suzaku} observations, it is clear that the properties 
of the pulsar such as emission mechanism, shape of pulse profiles observed during 
the {\it NuSTAR} observation are unique and had never been seen before. Phase-averaged
spectroscopy of {\it NuSTAR} and {\it Swift} observations showed that 
several continuum models fitted the broad-band data very well with comparable values of
reduced chi-sqaure (see Table~\ref{table2}). To investigate the cause of peculiar 
features observed during the {\it NuSTAR} observation and the nature 
of continuum models, we attempted to carry out phase-resolved spectroscopy of
the {\it NuSTAR} observation. For this, we extracted phase-resolved spectra for 25 phase 
bins by using {\tt XSELECT} package. With appropriate background, response and effective 
area files, phase-resolved spectroscopy was performed in 3--70 keV energy range. We 
fitted the phase-sliced spectra with several continuum models such as (i) Cutoff+BB, (ii) 
CompTT+BB and (iii) CompTT+Br along with photoelectric absorption component and a Gaussian 
function for $\sim$6.4~keV iron emission line. We was found that these three continuum  
models described phase-resolved spectra from {\it NuSTAR} observation well, yielding a 
reduced~$\chi^2$ close to 1 in each case. As the values of equivalent hydrogen column 
density and iron line width did not show any variation over pulse phase, these parameters 
were kept fixed at values obtained from phase-averaged spectroscopy (Table~\ref{table2}). 
In addition to these, we did not see any significant absorption like 
feature in the phase-sliced spectra. 

\begin{figure}
\centering
\includegraphics[height=2.8in, width=5.0in, angle=-90]{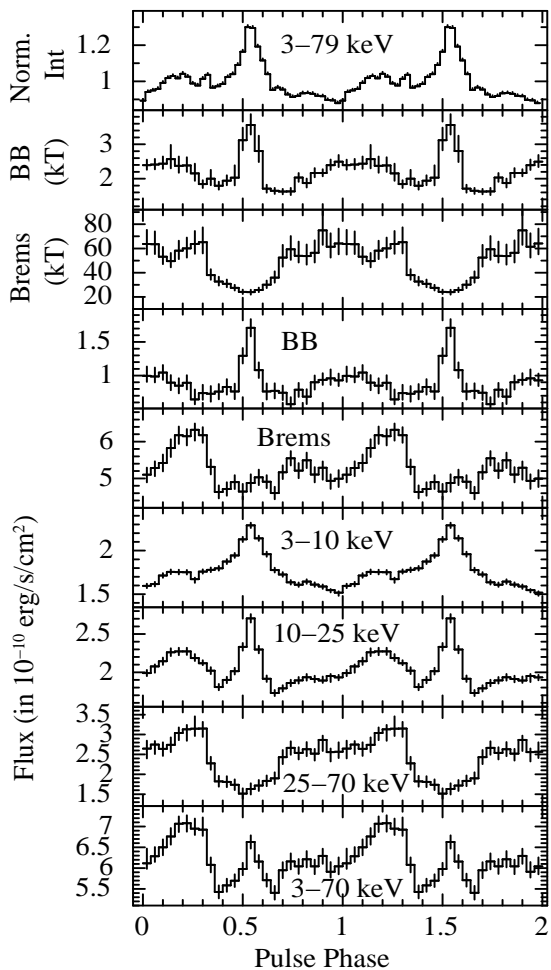}
\caption{Spectral parameters obtained from the phase-resolved spectroscopy 
of GX~1+4 during {\it NuSTAR} observation in October 2015 with Br+BB 
continuum model. Top panel shows the pulse profile of the pulsar in 3-79 keV range.
The parameters such as blackbody temperature, bremsstrahlung 
temperature, blackbody flux and bremsstrahlung flux (in unit of
10$^{-10}$~erg cm$^{-2}$ s$^{-1}$) are shown in the second, 
third, fourth and fifth panels, respectively. The source fluxes 
in 3-10 keV, 10-25 keV, 25-70 keV and 3-70~keV are given in sixth, 
seventh, eighth, ninth panels, respectively. The errors in the spectral 
parameters are estimated for 90\% confidence level.}
\label{fig7}
\end{figure}

Phase-resolved spectroscopy of {\it NuSTAR} observation, though yielded 
acceptable fits for all phase bins with above three continuum models, most of the 
spectral parameters did not show any significant variation (within errors) over pulse 
phases except the flux of thermal and non-thermal components. The flux of thermal 
components (blackbody and bremsstrahlung components) showed enhanced values in 0.5--0.6
phase ranges where a sharp peak was seen in the pulse profile below $\sim$25 keV (see
Fig.~\ref{fig3}). However, the flux of non-thermal component showed a shallow peak in 
0.0--0.3 pulse phase range. The shape of total flux profile did not match with 
that of the pulse profile of the pulsar which is possibly because of the significant 
flux differences in the soft and hard X-ray bands. As the flux of thermal (blackbody 
or bremsstrahlung) component is low by a factor of $\sim$10-20 as compared to the flux 
of non-thermal components (power law or CompTT; see Table~\ref{table2}), the hard 
X-ray flux dominated the soft X-ray flux leading to different shape of profile of 
total flux to that of the pulse profile. Apart from the shape of thermal, non-thermal 
and total flux profiles over pulse phase, the value of temperature of thermal component 
was also relatively high in 0.5--0.6 phase bin. Therefore, it is now confirmed that the 
narrow peak in the pulse profile below $\sim$25 keV was due to a thermal component which 
dominated emission in soft X-rays. Other than this, phase-resolved spectroscopy of 
{\it NuSTAR} observation of the pulsar with (i) Cutoff+BB, (ii) CompTT+BB and (iii) 
CompTT+Br continuum models did not provide any other information in understanding 
the properties of the source during this peculiar observation. 

Following this, we fitted the phase-resolved spectra obtained from the 
{\it NuSTAR} observation of the pulsar with Br+BB continuum model. In contrast 
to the results obtained from the phase-resolved spectroscopy with other three 
models (described earlier in this section), the Br+BB continuum model produced 
a better constraint on the spectral parameters. The parameters obtained from 
fitting all 25 phase-resolved spectra with this model are shown in Fig.~\ref{fig7} 
along with the pulse profile of the pulsar in the top panel. From the figure, it 
can be seen that the blackbody temperature and its flux were changing significantly 
with pulse phase and peaking in the same phase range (0.5--0.6 phase) at which the 
sharp peak was detected in profiles below $\sim$25 keV. The temperature and flux of 
the bremsstrahlung component were also found to be strongly variable across pulse 
phases and followed the shape of hard X-ray pulse profiles at energies $>$30~keV 
(Fig.~\ref{fig3}). The unabsorbed flux in different energy bands such as 3--10 keV, 
10--25 keV, 25--70 keV and 3--70 keV ranges, estimated from phase-resolved spectroscopy, 
are shown in sixth, seventh, eighth and ninth panels of the figure, respectively. 
The fluxes in these bands were found to be consistent with the pulse profiles in 
same energy ranges. Based on comparative studies of spectral models, we found that 
the Br+BB continuum model successfully described the phase-averaged and phase-resolved 
spectra of GX~1+4 better than all other models. This also provides an important clue 
on the peculiar evolution of pulse profiles and its corresponding spectral components 
during the {\it NuSTAR} observation in a faint state of the pulsar.

\section {Discussion}

\subsection{The origin of the peculiar pulse profile at low intensity level of GX~1+4}

Understanding the pulse profiles of accretion-powered X-ray pulsars illustrates 
the characteristics of the emission geometry and its beam pattern originating from 
the accretion column. Generally, these profiles are expected to be smooth and 
single peaked. However, most of the transient Be/X-ray binary pulsars such as 
EXO~2030+375 and GX~304-1 (\citealt{Naik2013, Epili2017, Jaisawal2016}) show
profiles with complicated structures including multiple dips or notches at 
various pulse phases. These features are found to be strongly energy dependent 
and seen up to high energies in EXO~2030+375 \citep{Naik2013}. Using phase-resolved 
spectroscopy, a high value of column density was estimated at these phases of pulse 
profiles. This suggested that the dips in the pulse profiles of Be/X-ray binary pulsars 
are due to absorption of X-ray photons by dense streams of matter that are
phase-locked with the neutron star.  

In the case of GX~1+4, the pulse profile is observed to be broad and singly peaked, 
as seen during the {\it RXTE} observations in 1996 to 2003 across a wide range of 
intensity (Fig.~\ref{fig2} of present work). Energy and luminosity dependence of
the pulse profiles have also been reported earlier (\citealt{Cui2004, Naik2005, 
Yoshida2017}). In addition to the broad shape, a characteristic sharp dip is also 
detected in the profiles at low and intermediate intensity levels. This feature was 
first detected in the data obtained from the {\it Ginga} observation of the pulsar 
\citep{Dotani1989}. The origin of this dip was thought to be related to the effect 
of cyclotron resonant scattering of photons near the neutron star surface. 
However, results from the {\it RXTE} and {\it Suzaku} observations suggest 
that this feature is associated with the obscuration or eclipse of emission 
region by the accretion column (\citealt{Giles2000, Galloway2000, Galloway2001, 
Yoshida2017}). In the present work, we have explored long term evolution of pulse 
profiles of the pulsar to understand the properties of the emission geometry at various 
luminosities. For the first time, unusual energy dependent peaks are detected in 
the pulse profiles of GX~1+4 during the {\it NuSTAR} observation. The observed profile 
is found to be strongly energy dependent. It shows a remarkable energy evolution by 
changing the broad shaped profile into a strong narrow peak below $\sim$25~keV. This 
peculiar peak disappears from the profile at higher energies. Moreover, another 
component also gradually emerges in the profiles at $\ge$15~keV. This leads to 
the origin of broad profiles at hard X-rays.

From the energy evolution, it can be considered that two emission components 
are contributing to the shape of pulse profiles during the {\it NuSTAR} observation. 
Corresponding spectral studies also confirm the presence of two components, 
blackbody and bremsstrahlung, in the broad-band energy spectrum of GX~1+4 
(see discussion below). With the help of phase-resolved spectroscopy, 
we find that the emission from blackbody component is peaking in a narrow 
pulse phase range. Therefore, it is possible that a hot emitting region is 
present in this phase causing the peculiar sharp peak in the profile. The 
broad-band continuum, expressed as a bremsstrahlung continuum or resulting 
from bulk Comptonization, is found to dominate in hard X-rays, producing 
a simple profile at higher energies.

\subsection{Spectroscopy}

The broadband energy spectrum of GX~1+4 has been described with standard 
models such as high energy cutoff power law, exponential cutoff power law 
and Comptonization of soft photons in a hot plasma, despite the complex 
processes occurring in the accretion column near the poles of the neutron 
star (\citealt{Galloway2000, Galloway2001, Cui2004, Naik2005, Ferrigno2007, 
Serim2017}). An additional blackbody component was also needed to describe the 
spectrum of the pulsar during high and intermediate intensity states 
(\citealt{Yoshida2017}). It is believed that the observed hard X-ray 
photons from accretion powered X-ray pulsars are due to the Comptonization 
of seed photons (blackbody, bremsstrahlung and cyclotron radiations) by 
the energetic electrons in the accretion column \citep{Becker2007}. 
The effects of bulk Comptonization as well as thermal Comptonization can 
lead to a power law spectrum with high exponential cutoff in these sources
(\citealt{Becker2007, Farinelli2012}).    

For investigating the spectral characteristics, physics of accretion column 
and the causes of peculiar peaks in the pulse profile as seen during the {\it NuSTAR}
observation, we have comprehensively studied the spectral properties of GX~1+4 
by using {\it RXTE}, {\it Suzaku} and {\it NuSTAR} observations over a wide range 
of luminosity. The 1-79 keV spectra from {\it NuSTAR} and {\it Swift} are well 
described by several traditional models along with an additional blackbody 
component. We find that a model consisting of a bremsstrahlung component
and a blackbody component (Br+BB) best describes the energy spectrum of GX~1+4
obtained from {\it NuSTAR} and {\it Swift} observations. This model provides a 
better understanding of the origin of the peculiar shape of pulse profile  
during the {\it NuSTAR} observation. Using the value of blackbody normalization
from spectral fitting and a source distance of 10 kpc, the radius of blackbody 
emitting region can be estimated to be $\approx$400~m. This indicates that the 
emitting region was located close to the neutron star surface or the base of the
accretion column.

Apart from the standard empirical models, physical models such as COMPMAG and BW 
models are also used to describe the spectra of GX~1+4 obtained from {\it NuSTAR} and 
{\it Swift} observations to understand the properties of the accretion column. These 
models were developed by considering the effects of Comptonization processes 
occuring near the neutron star surface (\citealt{Becker2007, Ferrigno2009, 
Farinelli2012}). Considering a distance of 4.3 and 10~kpc, the source 
luminosity was estimated to be $\approx$1.4 and 7.4$\times$10$^{36}$ erg s$^{-1}$, 
respectively. The critical luminosity in some of the pulsar sources e.g. 
EXO~2030+375, V~0332+53, GX 304-1 (\citealt{Mushtukov2015, Epili2017})
 is in the range of 10$^{37}$ erg s$^{-1}$, though at
sub-critical luminosity level, the 1-79 keV spectrum of GX~1+4 was successfully 
described with the BW model for bright sources. The physical parameters obtained 
from this model show interesting results. The ratio of bulk to thermal Comptonization 
$\delta$ is found to be relatively high by an order of magnitude during the 
observation. This indicates that the Comptonization from bulk motion of electrons 
with seed photons dominates the pulsar emission during a faint state. Similar 
outcome is also manifested from the COMPMAG2 model in which a lower electron temperature 
is detected at high terminal velocity $\beta$. \citet{Becker2005} have reported the 
dominance of bulk motion Comptonization in sub-critical pulsar (accreting below a 
specific luminosity), as observed in the present study. To investigate the 
changes in the emission process further, the BW model is used to describe
the spectra from a {\it Suzaku} observation at high intensity. 
We find that the effect of thermal Comptonization is high or relative 
close to the bulk emission process during bright state. However, 
we also noticed in this case that the parameter $\xi$ significantly differs 
from the condition $\xi$=2/$\sqrt{3}$ for the accretion flow that goes through  
a radiation dominated shock and satisfies the zero velocity condition 
at the neutron star surface (see Equation~26 from \citealt{Becker2007}).  
Advanced BW model is, therefore, needed to comprehend the nature of 
the X-ray continuum for sub-critical pulsars as observed with {\it Suzaku}. 

The BW model also provides opportunity to constrain the magnetic field of 
the neutron star. Several attempts have been made for estimating the field 
strength of GX~1+4 by using the standard accretion disk model as well 
as cyclotron resonance scattering feature (\citealt{Dotani1989, Makishima1988, 
Ferrigno2007}). It is interesting to mention that a high value of magnetic 
field, in the order of $\sim$10$^{13}$--10$^{14}$~G, was predicated by studying 
the spin-period evolution of the pulsar (\citealt{Dotani1989, Makishima1988, 
Cui2004}). A relatively lower value of field strength ($\sim$3$\times$10$^{12}$~G) 
was also reported by tentative detection of a cyclotron line at $\sim$34~keV 
(\citealt{Naik2005, Rea2005, Ferrigno2007}). Cyclotron lines are absorption 
like features observed in the hard X-ray spectrum of pulsars. These features 
are due to resonant scattering of photons with electron in a strong 
magnetic field of the order of 10$^{12}$~G \citep{Jaisawal2017}. Detection of 
these features, therefore, provides a direct method to estimate the  magnetic 
field of neutron stars. This line was tentatively reported at $\sim$34~keV in 
GX~1+4 using {\it BeppoSAX} and {\it INTEGRAL} observations. However, we do 
not find any evidence of this feature in the broad-band spectra of the pulsar 
obtained from {\it Suzaku} and {\it NuSTAR} observations. Using the BW model,
we put a tentative constrain on the magnetic field strength of the pulsar at 
$\sim$(5--10)$\times$10$^{12}$~G. Future observations in bright phase can 
possibly confirm this finding through the detection of cyclotron line in the 
pulsar spectrum.

\section {Summary and Conclusion}

In summary, a detailed timing and spectral study of accretion-powered
pulsar GX~1+4 was presented in this paper by using data from {\it RXTE},
{\it Suzaku}, {\it NuSTAR} and {\it Swift} observations. Our long term 
studies on pulse profile showed that the emission geometry of the pulsar is 
relatively simple, producing sinusoidal-like profiles that do not strongly 
depend on luminosity. During the {\it NuSTAR} observation, a peculiar narrow 
peak is detected in the profile below $\sim$25~keV. This peak is found 
to be strongly energy dependent. However, the hard X-ray profiles 
appear to be simple and broad peaked. The energy spectrum from {\it NuSTAR} 
is well described with a two component model consisting of a bremsstrahlung
and blackbody component along with iron fluorescence lines at $\sim$6.4 and 
6.9~keV. In the phase-resolved spectroscopy, a strong blackbody component is 
detected, peaking in a narrow pulse phase range of the pulsar and is interpreted 
as the cause of the peculiar peak in the profile. The BW and COMPMAG models are also 
used to explain the spectra of GX~1+4. We found that the effect of bulk 
Comptonization dominates the energy spectra in faint state. Though, we do not 
detect previously-reported cyclotron line in our study, the magnetic field 
is constrained to a range of $\sim$(5--10)$\times$10$^{12}$~G by using the BW model.

\section*{Acknowledgments}
We sincerely thank the referee for his/her suggestions on the paper.
GKJ thanks C. Ferrigno and M. Wolff for discussions on the physical models. 
The research work at Physical Research Laboratory is funded by the Department 
of Space, Government of India. The research leading to these results has 
received funding from the European Union's Seventh Framework Programme and 
Horizon 2020 Research and Innovation Programme under the Marie 
Sk{\l}odowska-Curie Actions grant no. 609405 (FP7) and 713683 (H2020; 
COFUNDPostdocDTU). This research has made use of data obtained through 
HEASARC Online Service, provided by the NASA/GSFC, in support of NASA 
High Energy Astrophysics Programs.


\label{lastpage}
\end{document}